\documentclass[journal]{IEEEtran}
\usepackage{amsmath, amscd, amsfonts, amssymb, graphicx, color}
\usepackage{array}
\usepackage{mathrsfs}
\usepackage[caption=false,font=normalsize,labelfont=sf,textfont=sf]{subfig}
\usepackage{textcomp}
\usepackage{stfloats}
\usepackage{url}
\usepackage{verbatim}
\usepackage{cite}
\usepackage{hyperref}
\hypersetup{
colorlinks=true,
linkcolor=blue,
citecolor=blue,
urlcolor=black
}

\usepackage{multirow}
\usepackage{multicol}
\usepackage{mathtools}
\usepackage{xr}
\usepackage{subfig}
\usepackage[flushleft]{threeparttable}
\usepackage{diagbox}
\usepackage{algorithm}
\usepackage{algpseudocode}
\usepackage[nameinlink,capitalise,noabbrev]{cleveref}
\crefname{figure}{Fig.}{Figs.}
\crefname{equation}{}{}
\usepackage{accents}

\newtheorem{theorem}{Theorem}
\newtheorem{corollary}{Corollary}
\newtheorem{proposition}{Proposition}
\newtheorem{lemma}{Lemma}

\newtheorem{definition}{Definition}
\newtheorem{example}{Example}

\newcounter{function}
\makeatletter
\newenvironment{function}[1][htb]{%
  \let\c@algorithm\c@function
  \renewcommand{\ALG@name}{Function}
  \begin{algorithm}[#1]%
  }{\end{algorithm}
}
\makeatother
\crefname{function}{Function}{Functions}
\Crefname{function}{Function}{Functions}

\begin{document}

\title{Unbiased and Error-Detecting Combinatorial Pooling Experiments with Balanced Constant-Weight Gray Codes for Consecutive Positives Detection}

\author{Guanchen He*, Vasilisa A. Kovaleva*, Carl Barton, Paul G. Thomas,
Mikhail V. Pogorelyy, Hannah V. Meyer$^{\#}$, and Qin Huang$^{\#}$
\thanks{These authors contributed equally: *, This work was supported by the National Natural Science Foundation of China under Grant 62071026 and 62331002. This work was also supported by the Simons Center for Quantitative Biology at Cold Spring Harbor Laboratory; US National Institutes of Health Grant S10OD028632-01. The funders had no role in the template design or decision to publish. (\textit{Corresponding authors \#: Qin Huang, Hannah V. Meyer})}
\thanks{Guanchen He and Qin Huang are with the School of Electronic and Information Engineering, Beihang University, Beijing 100191, China (e-mail: hgc01@buaa.edu.cn; qhuang.smash@gmail.com).}
\thanks{Vasilisa A. Kovaleva and Hannah V. Meyer are with the Simons Center for Quantitative Biology, Cold Spring Harbor Laboratory, NY 11724, USA (e-mail: kovaleva@cshl.edu; hmeyer@cshl.edu).}
\thanks{Carl Barton is with Birkbeck University of London, Malet Street, London, WC1E 7HX.}
\thanks{Mikhail V. Pogorelyy and Paul G. Thomas are with the Department of Host-Microbe Interaction, St. Jude Children’s Research Hospital, Memphis, TN 38105, USA.}
}

\maketitle
\begin{abstract}
Combinatorial pooling schemes have enabled the measurement of thousands of experiments in a small number of reactions. This efficiency is achieved by distributing the items to be measured across multiple reaction units called pools. However, current methods for the design of pooling schemes do not adequately address the need for balanced item distribution across pools, a property particularly important for biological applications.
Here, we introduce balanced \emph{constant-weight Gray codes for detecting consecutive positives} (DCP-CWGCs) for the efficient construction of combinatorial pooling schemes. Balanced DCP-CWGCs ensure uniform item distribution across pools, allow for the identification of consecutive positive items such as overlapping biological sequences, and enable error detection by keeping the number of tests on individual and consecutive positive items constant. 
For the efficient construction of balanced DCP-CWGCs, we have released an open-source python package \textit{codePub}, with implementations of the two core algorithms: a branch-and-bound algorithm (BBA) and a recursive combination with BBA (rcBBA). Simulations using \textit{codePub} show that our algorithms can construct long, balanced DCP-CWGCs that allow for error detection in tractable runtime.
\end{abstract}

\begin{IEEEkeywords}
Balanced constant-weight Gray codes, combinatorial pooling, consecutive positives detection, branch-and-bound algorithm, recursive combination.
\end{IEEEkeywords}

\section{Introduction}
\label{intro}

\begin{figure}[tbp]
    \centering
    \includegraphics[scale=0.58]{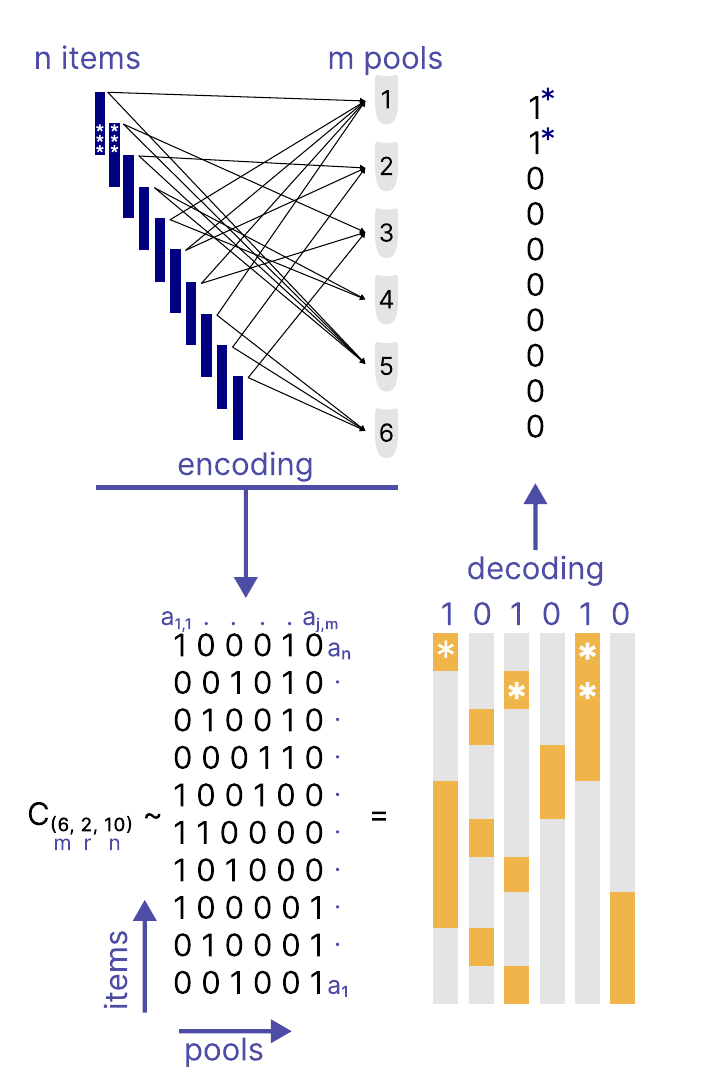}
    \caption{\textbf{A combinatorial pooling example with $10$ items and $6$ pools}. Items are mixed into pools according to an encoding based on a DCP-CWGC, represented as its incidence matrix $H$ and color-coded by the expected experimental outcome (grey-negative, yellow-positive). Knowing the DCP-CWGC, the experimental outcome can be decoded and the item of interest, as indicated by stars, identified. } 
    \label{fig:pooling_example}
\end{figure}

High-throughput experiments have become a staple across many domains of biomedical research. In these experiments, the aim is to obtain measurements for hundreds or thousands of samples in a single experimental setup. To reduce the experimental complexity and associated cost, combinatorial pooling strategies have been developed. In general terms, a combinatorial pooling experiment uses an encoding scheme according to which items (samples) are mixed such that each item is included across multiple pools and each pool contains multiple items. The experimental read-out happens on pool level, where the measurements across pools yield a unique pattern of signals. This signal is then decoded to identify the individual-level item measurements (\cref{fig:pooling_example}).  
Since the number of pools is typically much smaller than the total number of items, combinatorial pooling offers better efficiency than individual testing. In addition, well-designed encoding/decoding schemes will allow for experimental error detection by taking advantage of items being present across multiple pools.

\begin{figure}[htbp]
\centering
\includegraphics[scale=0.5]{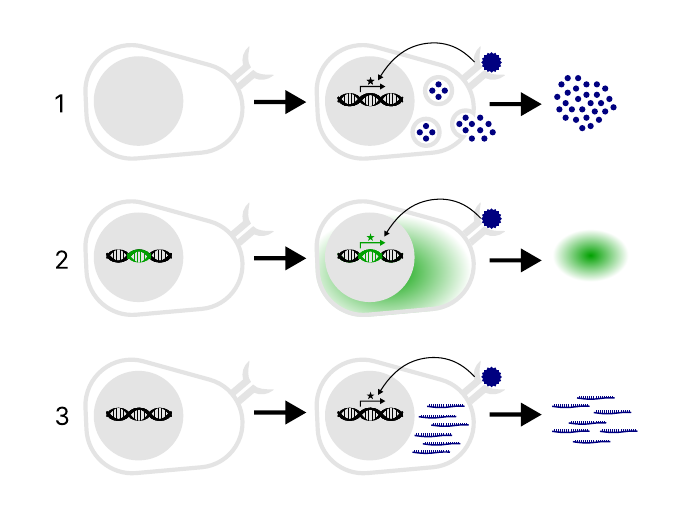}
\caption{\textbf{Experimental assays used in combinatorial pooling experiments.} In each example, a response can be observed upon successful binding of the item (blue circle) to a cell surface receptor. The item(s) that yield such a response are considered positive. Examples of responses and their read-outs are: 1. A primary cell line which secretes activation markers such as cytokines, quantifiable by biochemical assays, 2. a reporter cell line, which produces a fluorescently labeled protein quantifiable by e.g. flow cytometry or microscopy 3. gene expression changes in primary cells, quantifiable by RNA-sequencing.  } 
\label{fig:experiment}
\end{figure}

For biological applications, the total number of pools, items per pool and the number of tests per item are key experimental parameters to consider. 
Minimizing both the number of pools and items per pool is crucial in biological assays with limited input material, for instance when assessing the response of patient-derived, primary cells to stimuli (items) by measuring the secretion of activation markers (\cref{fig:experiment}-1). In cases where the input material is less constrained, such as using a reporter cell line (\cref{fig:experiment}-2), maintaining a constant number of tests per item is crucial to prevent measuring bias in the experiment. 
All these factors intersect in experiments where experimental conditions and read-out are constrained by volume and range, such as measuring gene expression in a heterogeneous cell population using RNA sequencing (\cref{fig:experiment}-3). Therefore, 
the experimental design should enable adjusting parameters to minimize the number of pools and items per pool, while also ensuring a constant number of tests per item for balanced testing. Designing a pooling scheme that adheres to these constraints and allows for error correction is non-trivial. This is further compounded when the items to be tested are non-independent, a common case in biological pooling experiments. For instance, in schemes designed to detect positive items within a pool of protein sequences, sequences are often overlapping, and these consecutive sequences will yield non-independent positive signals.

Here, we propose balanced \emph{constant-weight Gray codes for detecting consecutive positives} (DCP-CWGCs) as a new combinatorial pooling scheme addressing these experimental considerations. Our scheme organizes items into pools following a structured sequence called DCP-CWGC, which consists of distinct binary vectors (binary addresses of items) indicating specific pools into which items are mixed. Binary addresses have a constant Hamming weight, while adjacent addresses have a Hamming distance $2$ and unique OR-sums. These DCP-CWGC-defining constraints enable the identification of any consecutive positive items, keep the number of tests on each item and on each pair of consecutive items constant, and facilitate error detection. 
To ensure stable and unbiased detection results across pools, particular attention is placed on DCP-CWGCs that produce \emph{balanced} pooling arrangements, where the number of items per pool is approximately constant across all pools. 

To construct balanced DCP-CWGCs with flexible parameters, we developed a \emph{branch-and-bound algorithm} (BBA), which efficiently constructs DCP-CWGCs with near-optimal balance for codes with short and moderate lengths (3000 items in less than 250 seconds).
BBA conducts a depth-first heuristic search for a balance-optimized path in the bipartite graph formed from the addresses of items and the unions of consecutive addresses. 
A balance-optimized path is a path through the bipartite graph where, when looking at the address nodes on the path, each pool occurs an approximate equal number of times.
Furthermore, we show that a long code can be constructed by a recursive combination of several short codes. We prove that for a range of parameter combinations, the recursive approach achieves maximal lengths and perfect balance. To broaden the applicability outside this parameter range, we implemented an extension of BBA that constructs long codes by recursive combination of several short, BBA-generated codes. This \emph{recursive combination together with branch-and-bound algorithm} (rcBBA) further extends achievable code lengths in limited runtime, while maintaining a near-perfect balance. 
BBA and rcBBA are both implemented in an open-source software \textit{codePub}, and simulation results show that they construct balanced DCP-CWGCs for thousands of items in tractable time.

Our balanced DCP-CWGC scheme will find applications across many branches of biological research. Its ability to ensure a constant number of tests on consecutive items makes it particularly useful for protein and DNA pooling experiments, where consecutive items represent overlapping amino acid or nucleotide sequences. 
For instance, a common effort in adaptive immunology research is identifying the epitopes that T cell and B cell receptors are reactive against. These epitopes are short peptide sequences generated from cellular proteins and their identification out of the large pool of possible peptides can be aided by combinatorial pooling experiments \cite{sospedra2003use,klinger_multiplex_2015,nolan_large-scale_2025}, where DCP-CWGC-based experimental designs can greatly reduce the experimental complexity. In plants, CLE peptides are critical signaling molecules for cell division and differentiation.  These peptides are generated from long precursor proteins and a DCP-CWGC-based combinatorial pooling experiment could aid in systematic screens to locate mature CLE peptides within these  precursors \cite{gao2012cle}. New methods developed for single-cell genomics have enabled large-scale genetic screening studies, such as the genome-wide identification of genes and their regulatory regions \cite{gasperini_genome-wide_2019,alda2024mapping}. While these approaches have already implemented highly multiplexed screening methods, they were not able to comprehensively assess the regulatory DNA sequences surrounding each gene. DCP-CWGC-based combinatorial pooling experiments could facilitate a comprehensive survey by densely tiling candidate DNA regions of interest.

In the remainder of this paper we will introduce prior work on combinatorial pooling designs and combinatorial Gray codes for consecutive positives detection (\cref{sec:preliminaries}); define DCP-CWGCs and balanced DCP-CWGCs (\cref{sec:DCP-CWGC}); describe BBA and the recursive combination approach (\cref{sec:BBA,sec:RCA}, respectively); present the computational complexity analysis and simulation results of BBA and rcBBA (\cref{sec:results}) and, lastly, summarize our methods and results (\cref{sec:conclusion}).

\section{Preliminaries}
\label{sec:preliminaries}

\subsection{Combinatorial Pooling Designs}
Consider there are $n$ items $X=\{x_1,x_2,\cdots,x_n\}$ and $m$ pools $Q=\{q_1,q_2,\cdots,q_m\}$. A combinatorial pooling design is characterized by a code comprising $n$ binary vectors, $C=\{a_1,a_2,\cdots,a_n\}$, where each vector $a_j=(a_{j,1},a_{j,2},\cdots,a_{j,m})$, $j=1,\cdots,n$, is called the (binary) address of the item $x_j$. For $1 \leq j \leq n, 1 \leq i \leq m$, $a_{j,i} = 1$ indicates that the $j$-th item $x_j$ is contained in the $i$-th pool $q_i$; $a_{j,i} = 0$ otherwise. For the unique interpretation of each item, all addresses in the code should be different. This code is identical to an $m \times n$ \emph{incidence matrix} $H=(a_{j,i})_{1 \leq i \leq m, 1 \leq j \leq n}$ over a binary Galois field $\text{GF}(2)$, where each column in $H$ is the transpose of a binary address in the code (\cref{fig:pooling_example}, lower panel). The columns of $H$ correspond to the items, and the rows of $H$ correspond to the pools. As long as there is no ambiguity, hereafter we use the concept of code for a combinatorial pooling design and its corresponding incidence matrix interchangeably.

Combinatorial pooling design requirements differ from application to application and have thus motivated the construction of codes with specific properties, such as disjunctness and separability. For a comprehensive survey, the reader is referred to \cite{Du2006PoolingDA,Du2000CombinatorialGT}.
Here, we will focus on a special class
that takes the shape of combinatorial Gray codes \cite{savage_survey_1997,mütze2024combinatorialgraycodesanupdated} for consecutive positives detection, where each binary address differs from its consecutive ones by a ``small'' change.

\subsection{Combinatorial Gray Codes for  Consecutive Positives Detection}
\begin{figure}[htbp]
\centering
\includegraphics[width=0.3\textwidth]{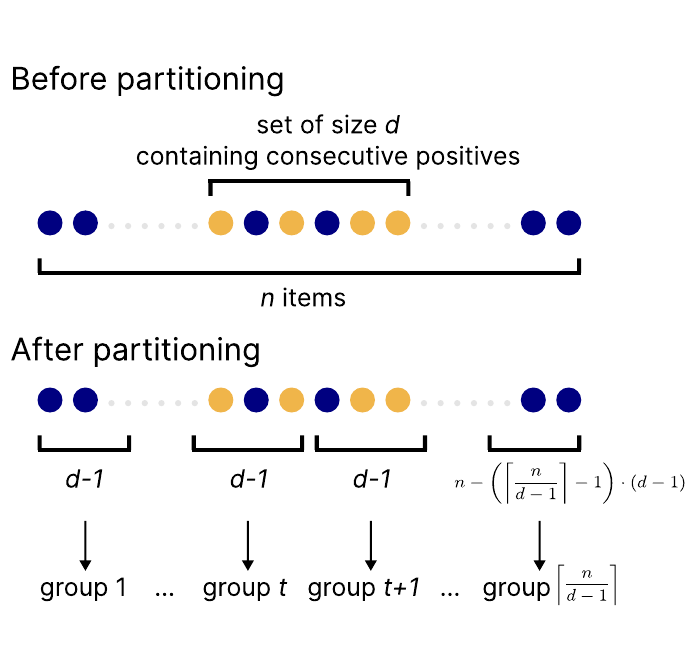}
\caption{\textbf{Combinatorial Gray codes for the
design of combinatorial pooling with consecutive positives
detection.} The uniform partition reduces the general $d$ consecutive positive case to the $2$ consecutive positive case. The yellow circles represent positive items and the blue circles represent negative items.}
\label{fig:partition}
\end{figure}

Colbourn introduced combinatorial Gray codes for the design of combinatorial pooling experiments with consecutive positives detection \cite{colbourn_group_1999}. He considered a general case where $n$ items to be tested are linearly ordered and positive items are within a consecutive set of size at most $d$. He illustrated that by uniformly partitioning the $n$ linearly ordered items into $\lceil \frac{n}{d-1} \rceil$ groups, the general case of detecting $d \geq 2$ consecutive positive items can be simplified to the case of $d=2$. As shown in \cref{fig:partition}, after partitioning, there are at most $2$ consecutive groups containing positive items. 
If there exists a code $C$ of length $\lceil \frac{n}{d-1} \rceil$, where any address and any bitwise OR-sums of consecutive addresses are pairwise different, the corresponding combinatorial pooling design can distinguish up to $2$ consecutive positives. 
While optimized for the number of pools, this scheme only resolves when no experimental testing error occurred and can thus not address false positive or negative outcomes.

A number of code designs specifically address error detection and correction for combinatorial pooling experiments with consecutive positive detection.
While different in construction, these codes are all equivalent to sequences comprising distinct binary addresses. In the first of a series of papers, Muller and Jimbo introduced $m \times n$ $2$-consecutive positive detectable matrices with fixed constant column weight $r$ \cite{muller_consecutive_2004}. They proved the existence of maximal $2$-consecutive positive detectable matrices with $n=\binom{m}{r}$ for any $m \in \mathbb{N}$ and $r, 1 \leq r \leq \lfloor \frac{m}{2} \rfloor$. In follow up work\cite{muller_cyclic_2008}, they recursively construct cyclic sequences of all $r$-subsets of $\{1,2,\cdots,m\}$ with distinct consecutive unions, in which these unions are either all of even order or all of odd order. These sequences enable the detection of up to $1$ error, and are proven to exist for $r=2,3,\cdots,7$ and sufficiently large $n$. Orthogonal construction approaches use block sequences of maximal $t$-packings, where all blocks and all unions of two consecutive blocks consist of an error correcting code with minimum distance $4$. Such sequences can not only identify any $2$ consecutive positives but also correct a single error\cite{momihara_constructions_2008,ge_block_2009}.

\begin{figure}[htbp]
\centering
\includegraphics[width=0.35\textwidth]{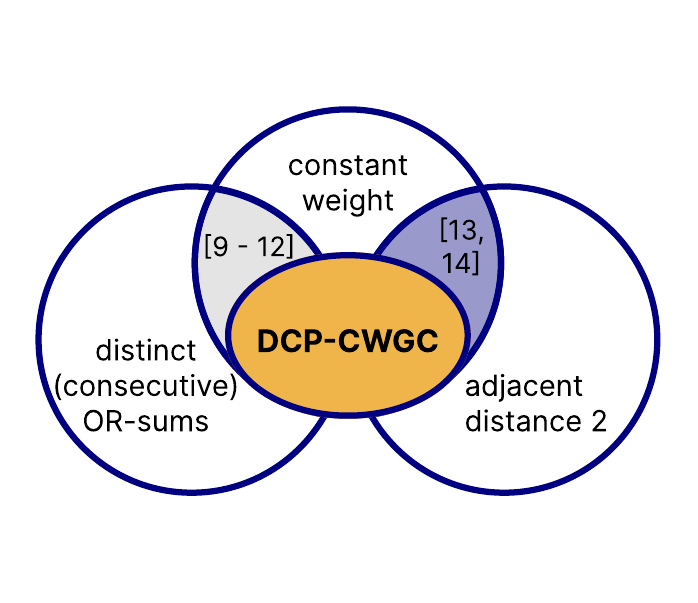}
\caption{\textbf{Properties of existing coding schemes.} DCP-CWGCs bridge the gap between existing codes capable of consecutive positives detection (grey) and with constant weight of both codewords and consecutive OR-sums (blue).}
\label{fig:venn}
\end{figure}

All codes above \cite{muller_consecutive_2004,muller_cyclic_2008,momihara_constructions_2008,ge_block_2009} offer constant weight and distinct OR-sums, which ensure the unique identification of each pair of consecutive items (\cref{fig:venn}, gray). However, they are limited in the parameter range for which codes can be generated but cannot ensure balance property when shortened to flexible lengths. Moreover, in these codes, the OR-sums of consecutive addresses are not necessarily constant-weight, thus they do not necessarily provide an equal number of tests for each pair of consecutive items. 
Another set of codes, originally developed for local rank modulation in storage systems and A/D conversion in communication systems \cite{gad_constant-weight_2011,tang_distance-2_1973}, offer constant-weight and constant distance between adjacent codewords (\cref{fig:venn}, blue), yielding constant weight for OR-sums of consecutive codewords. However, they do not fulfill the OR-sums constraint and thereby cannot be applied to the consecutive positives detection. The synthesis of properties of these codes characterizes our proposed DCP-CWGCs.

\section{Balanced Constant-Weight Gray Codes for Detecting Consecutive Positives}
\label{sec:DCP-CWGC}

In this section, we first define DCP-CWGCs based on their key features. We then continue with a characterization of their balance property which ensures a consistent number of items across pools and therefore stable detection results across pools in biological combinatorial pooling designs.

\subsection{Definition of DCP-CWGCs}
\label{sub:dcp-cwgc}

\begin{definition}
    \label{def:dcp-cwgc}
    An $(m,r,n)$ DCP-CWGC is a sequence of distinct binary addresses $C=\{a_1,a_2,\cdots,a_n\}$, $a_j=(a_{j,1},a_{j,2},\cdots,a_{j,m}),j=1,\cdots,n$, such that the following three constraints are satisfied:
    \begin{enumerate}
    \item Distinct OR-sums constraint: $\forall j, k \in \{1, 2, \dots, n-1\}, \, j \neq k$,
    \begin{equation}
        \label{item:constraint1}
        a_{j} \vee a_{j+1} \neq a_k \vee a_{k+1},
    \end{equation}
    where "$\vee$" represents the bitwise OR-sum of two binary vectors.
    \item Constant-weight constraint: $\forall j \in \{1, 2, \dots, n\}$,
    \begin{equation}
        \label{item:constraint2}
        \sum_{i=1}^{m} a_{j,i}=r.
    \end{equation}
    \item Adjacent distance constraint: $\forall j \in \{1, 2, \dots, n-1\}$,
    \begin{equation}
        \label{item:constraint3}
      D_H(a_j,a_{j+1})=2,
    \end{equation}
    where ``$D_H$'' represents the Hamming distance of two binary vectors.
\end{enumerate}
\end{definition}

An $(m,r,n)$ DCP-CWGC directly corresponds to a pooling design with $m$ pools, the address weight $r$, and $n$ items. Constraint \cref{item:constraint1} ensures the unique identifier for each pair of consecutive items. Constraint \cref{item:constraint2} ensures the constant number of tests for each item. A combination of constraints \cref{item:constraint2} and \cref{item:constraint3} results in the constant weight equal to $r+1$ for the OR-sum of each pair of consecutive addresses, thus leading to a constant number of tests for each pair of consecutive items.
Therefore, as long as we know whether there is a single positive item or two consecutive positive items, we can always detect at least $1$ error based on the number of positive pools in the experiment.

Throughout the paper, we use the following notation.
We denote the ensemble of all $(m,r,n)$ DCP-CWGCs by $\text{DCP-CWGC}(m,r,n)$. The ensemble of all DCP-CWGCs with parameters $(m,r)$ and arbitrary length is denoted by $\text{DCP-CWGC}(m,r)$. We denote the set of all binary vectors of length $m$ and weight $r$ by $S(m,r)$. An address $a$ can be denoted as its binary form, e.g. $a=(1,1,0,0,0,1)$, or as a corresponding index set $I(a)=\{1,2,6\}$. Consequently, the bitwise OR-sum of two binary addresses is equivalent to the union of their index sets. Hereafter, as long as there is no ambiguity, we use ``unions of index sets of addresses'', ``unions of addresses'' and ``bitwise OR-sums of addresses'' interchangeably. 

In combinatorial pooling experiments, often it is preferred to use as low a number of pools, and items per pool as possible due to restrictions on input material available and on the sensitivity of the measuring method. Consequently, it is important to minimize the number of pools $m$ or the number of pools per item $r$ in the pooling scheme. 
On the other hand, if $m$ and $r$ are predetermined by experimental conditions, then it will be of interest to find the maximum number of items $n$ that can be tested. 
The next proposition gives an upper bound of $n$ with respect to $m$ and $r$ for a DCP-CWGC.\\

\begin{proposition}
\label{prop:bound}
  The length $n$ of any code in $\text{DCP-CWGC}(m,r)$ is upper bounded by 
  \begin{equation}
    \label{eq:bound}
      n \leq \min \left\{\binom{m}{r},\binom{m}{r+1}+1\right\}
  \end{equation}
\end{proposition}

\begin{IEEEproof}
    Consider a code $C=\{a_1,a_2,\cdots,a_n\} \in \text{DCP-CWGC}(m,r)$. The addresses in $C$ all come from the set $S(m,r)$, and they are pairwise distinct. Thus, $n \leq \binom{m}{r}$. The OR-sums of consecutive addresses in $C$ all come from the set $S(m,r+1)$, and they are pairwise distinct. Thus, $n-1 \leq \binom{m}{r+1}$.
\end{IEEEproof}

\begin{example}
    \label{ex:incidence_matrices}
    For the following incidence matrix 
    $$H_{5,2,10}=\left[\begin{array}{cccccccccc}
      0 & 1 & 1 & 0 & 0 & 0 & 0 & 0 & 1 & 1\\
      1 & 0 & 0 & 1 & 0 & 0 & 1 & 0 & 0 & 1\\
      1 & 1 & 0 & 0 & 1 & 1 & 0 & 0 & 0 & 0\\
      0 & 0 & 1 & 1 & 1 & 0 & 0 & 1 & 0 & 0\\
      0 & 0 & 0 & 0 & 0 & 1 & 1 & 1 & 1 & 0\\
    \end{array}\right],$$
    columns are pairwise distinct, and it satisfies all constraints in \cref{def:dcp-cwgc}. Moreover, the bound (\cref{prop:bound}) is met, thus it defines a $(5,2,10)$ DCP-CWGC that achieves maximal length. 
\end{example}

\subsection{Balanced DCP-CWGCs}

In balanced DCP-CWGCs, 
the sum of each row in their corresponding incidence matrices is approximately constant.

To characterize the balance property of a code $C \in \text{DCP-CWGC}(m,r,n)$, we define its balance vector as $W_{C}=(w_1,w_2,\cdots,w_m)$, where $$w_i=\sum_{j=1}^{n} a_{j,i}, 1\leq i \leq m.$$
The deviation of the balance vector, $\delta_{C} := \max(W_{C}) - \min(W_{C})$, is a critical evaluation criterion of the balance property of $C$. The smaller $\delta_{C}$, the better balanced the DCP-CWGC becomes, resulting in a more stable detection result for the corresponding combinatorial pooling arrangement. Since a code $C$ can be identically represented by its incidence matrix $H$, we also use $W_{H}$ and $\delta_H$ to denote the balance vector of $C$ and its deviation, respectively. 

The DCP-CWGC in \cref{ex:incidence_matrices} is strictly balanced, that is, its $\delta_{H_{5,2,10}} = 0$. In fact, any maximal $\left(m,r,n=\binom{m}{r}\right)$ DCP-CWGC is bound to be strictly balanced since all $\binom{m}{r}$ binary addresses occur exactly once. In\cref{sec:RCA} we show that there exist maximal $\left(m,r,n=\binom{m}{r}\right)$ DCP-CWGCs for all positive integers $r$ and $m\geq r+1$ that exhibit the perfect balance property.

\begin{proposition}
\label{prop:balance}
  An $\left(m,r,n =\binom{m}{r}\right)$ DCP-CWGC $C$ has a perfect balance vector $W_{C}=(w_1,\cdots,w_m)=\left(\binom{m-1}{r-1},\cdots,\binom{m-1}{r-1}\right)$.
\end{proposition}

Given arbitrary code parameters $(m,r,n)$, it is generally difficult to ensure perfect balance. For instance, if $n \cdot r \  \mathrm{mod}\ m \neq 0$, the deviation of the balance vector for a $(m,r,n)$ DCP-CWGC cannot be zero. Thus, in the following sections, we will present numerical algorithms that construct DCP-CWGCs exhibiting a nearly perfect balance property with flexible parameters.

\section{Branch-and-Bound Algorithm}
\label{sec:BBA}

In this section, we propose a branch-and-bound algorithm (BBA) to construct balanced DCP-CWGCs with flexible parameters $(m,r,n)$. BBA proceeds by depth-first heuristic search for a path in an address-union bipartite graph. While traversing this graph, BBA evaluates the balance property of all potential next nodes from the current state and selects the node that best maintains balance. Once the path reaches the target length, the algorithm terminates and returns the DCP-CWGC as the sequence of the addresses in the path.

\begin{figure}[htbp]
\centering
\includegraphics[width=0.4\textwidth]{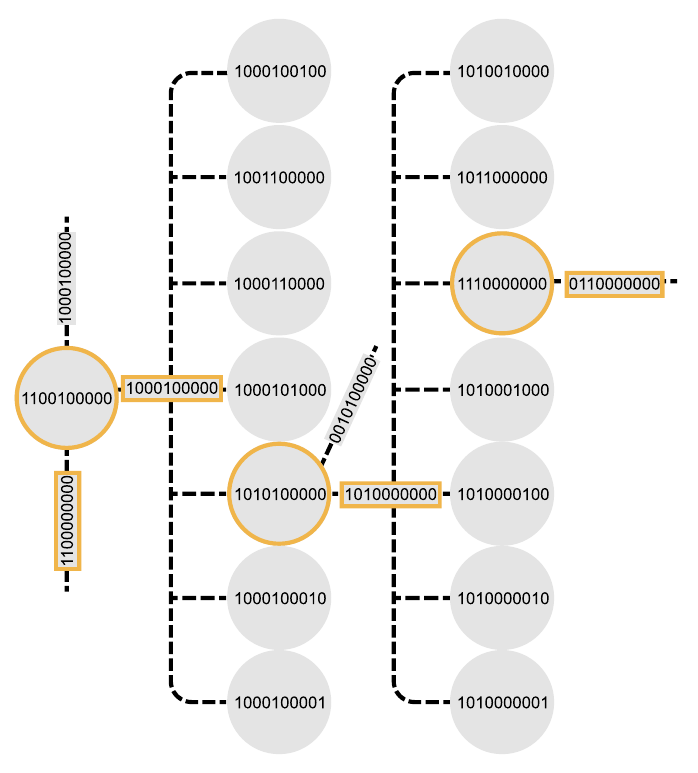}
\caption{\textbf{An example of the balance-optimized path search in the BBA scheme.} Rectangular and circular nodes represent addresses ($a$) and unions ($u$), respectively. Yellow nodes indicate the search path taken in which the rectangular ones correspond to a possible segment in an $(m=10,r=2,n \geq 4)$ DCP-CWGC.}
\label{fig:bba}
\end{figure}

An example of the BBA (\cref{alg:bba}) traversing the bipartite graph is presented in \cref{fig:bba}.
The nodes in the graph are divided into address nodes ($a$, rectangular nodes in \cref{fig:bba}) and union nodes ($u$, circular nodes in \cref{fig:bba}).
Given the constant weight for addresses and OR-sums of consecutive addresses, the set of all possible addresses $AN$ equals to $S(m,r)$, and the set of all possible unions $UN$ equals to $S(m,r+1)$. We define $Adj(u)$ as the subset of $AN$ incident to union $u$ such that bitwise OR-sum of $u$ and every address $a \in Adj(u)$ equals $u$: $$\forall a \in Adj(u): a \vee u = u.$$ Similarly, we define $Adj(a)$ as the subset of $UN$ incident to $a$ such that the bitwise OR-sum of $a$ and every union $u \in Adj(a)$ equals $u$: $$\forall u \in Adj(a): a \vee u = u.$$
We denote the path consisting only of address nodes as $A$ and the path consisting only of union nodes as $U$. 
BBA is initiated with an arbitrary address $a_1 \in AN$ and arbitrary union $u_1 \in UN$ such that $u_1 \in Adj(a_1)$.
Next, a balance-optimized path $A + U$ containing $n$ address nodes is determined heuristically, alternating between address nodes $a$ and union nodes $u$ and selecting nodes that maintain the balance best.

\begin{algorithm}[hbt!]
\renewcommand{\thealgorithm}{1}
\caption{Branch-and-bound Algorithm}\label{alg:bba}
\begin{algorithmic}[1]
\Require the parameters $(m,r,n)$ of DCP-CWGC, first address $a_1$, desired balance vector $W_{des}$.
\Ensure the constructed balanced DCP-CWGC $A$.
\State initiate the path of addresses $A=\{a_1\}$
\State initiate path of unions $U=\phi$.
\State randomly choose current union node $u_1$ from the subset $Adj(a_1)$.
\State $A=\texttt{SearchPath}(A, U, u_1, W_{des})$.
\State \Return $A$.
\end{algorithmic}
\end{algorithm}

Here, we outline the specific steps in the heuristic balance-optimized path search, with pseudocode for the core recursive function $\texttt{SearchPath}$ depicted in \cref{fun:search_path_bba}.
\begin{itemize}
    \item We initiate the path of the unions by $U=\phi$ and the path of the address $A$ by the address $a_1$; if $a_1$ is not specified, choose a random $a_1 \in AN$, $A=\{a_1\}$.
    \item We choose the next union $u_1$ (\cref{fun:search_path_bba}, lines 17-26) to be added to the path. For each $u \in Adj(a_1)$, we calculate the balance penalty score associated with adding $u$ to the path and choose the $u$ with minimal penalty as the next union $u_1$: $U \gets U \cup u_1 = \{u_1\}$.
    \item We choose the next address $a_2$ (\cref{fun:search_path_bba}, lines 1-16) to be added to $A$, in analogy to step two for finding the next union.  We update $A$ by adding $a_2$: $A \gets A \cup a_2 = \{a_1, a_2\}$  
    \item We proceed further as described in steps 2 and 3, alternating between updating $A$ and $U$. We stop when the length of $A$ reaches the desired length $n$. Then $A$ is returned as the constructed balanced DCP-CWGC.
\end{itemize}

The branch-and-bound algorithm sequentially extends $A$ and $U$ by choosing the nodes that best maintain balance.
However, sometimes it may reach a dead end, i.e., for node $p_i$, $Adj(p_i) \backslash (A \cup U)$ may be an empty set. Then, the algorithm needs to backtrack to the previous node $p_{i-1}$ (\cref{fun:search_path_bba}, lines 16 and 26) and choose a next node $p_j$ different from $p_i$. If such a node exists, then $p_i$ in $A$ or $U$ depending on the type of node is replaced by $p_j$, and the search for the balance-optimized pathway continues. Otherwise, the algorithm backtracks again.  To avoid these dead ends in the graph, \texttt{SearchPath} function operates recursively (\cref{fun:search_path_bba}, lines 11 and 24).

\begin{function}[hbt!]
\caption{\texttt{SearchPath}($A, U, p, W_{des}$)}
\label[function]{fun:search_path_bba}
\begin{algorithmic}[1]
\Require path of addresses $A$, path of unions $U$, current node $p$; the desired balance vector $W_{des}$.
\Ensure the found path of addresses $A$ such that $A.length = n$.
\If{$p \in AN$ and $p \not\in A$}
    \State $A = A \cup \{p\}$.
    \If{$A.length = n$}
        \State \Return $A$.
    \EndIf
    \For{$p_{next} \in Adj(p)$} 
    \State $penalty_{p_{next}} = variance(W_{des} - W_{A \cup p_{next}})$.
    \EndFor
    \State sort $Adj(p)$ using $penalty$ in ascending order $\to Adj(p)_{sorted}$.
    \For{$p_{next} \in Adj(p)_{sorted}$}
        \State $A_{updated}=\texttt{SearchPath}(A, U,p_{next},W_{des})$.
        \If{$A_{updated}.length = n$}
            \State \Return $A_{updated}$.
        \EndIf
    \EndFor
    \State $A = A \backslash \{p\}$.
\ElsIf{$p \in UN$ and $p \not\in U$}
    \State $U = U \cup \{p\}$.
    \For{$p_{next} \in Adj(p)$}
    \State $penalty_{p_{next}} = variance(W_{des} - W_{A \cup p_{next}})$.
    \EndFor
    \State sort $Adj(p)$ using $penalty$ in ascending order $\to Adj(p)_{sorted}$.
    \For{$p_{next} \in Adj(p)_{sorted}$}
        \State $U_{updated} = \texttt{SearchPath}(A, U, p_{next}, W_{des})$.
    \EndFor
    \State $U = U \backslash \{p\}$.
\EndIf
\end{algorithmic}
\end{function}

\section{Recursive Combination Approach}
\label{sec:RCA}

In this section, we describe our recursive combination approach. We prove that this approach can combine short DCP-CWGCs to construct  maximal strictly balanced $(m,r,n=\binom{m}{r})$ DCP-CWGCs for any positive integer $r$ and $m \geq r+1$. We then pair the recursive combination approach with BBA, and show that we can construct approximately balanced DCP-CWGCs with flexible parameters $(m,r,n)$. 

\subsection{Recursive Combination of DCP-CWGCs}

The recursive combination approach is based on a combination operation that concatenates the incidence matrices of two DCP-CWGCs. These two component codes are in the ``augmented'' form, whose incidence matrices have one row as an all-one vector $\mathbf{1}$ or an all-zero vector $\mathbf{0}$. We first introduce two useful augmentation operators for notational brevity, then demonstrate the details of the approach. 

For a binary vector $v=(v_1,v_2,\cdots,v_m)^T$, we denote its ``+'' augmented vector as $v^{+}=(v_1,v_2,\cdots,v_m,1)^T$, and we denote its ``-'' augmented vector as $v^{-}=(v_1,v_2,\cdots,v_m,0)^T$. For an $m \times n$ binary matrix $H=[a_1,a_2,\cdots,a_n]_{m \times n}$, we denote its ``+'' augmented matrix as
$
H^{+} :=\left[\begin{array}{cccccc}
    a_1 & a_2 & \cdots & a_{n}\\
    1 & 1 & \cdots & 1 
    \end{array}\right]_{(m+1)\times n}, 
$
and denote its ``-'' augmented matrix as 
$
H^{-} :=\left[\begin{array}{cccccc}
    a_1 & a_2 & \cdots & a_{n}\\
    0 & 0 & \cdots & 0 
    \end{array}\right]_{(m+1)\times n}.   
$
Moreover, for a DCP-CWGC $C$ whose incidence matrix is $H$, we denote $C^{+}$ as the augmented DCP-CWGC whose incidence matrix is $H^{+}$, and denote $C^{-}$ as the augmented DCP-CWGC whose incidence matrix is $H^{-}$. These two operators may be repeatedly applied. For example, $
v^{2+,-}=(v_1,\cdots,v_m 1,1,0)^T$, or
$$
H^{+,2-}=\left[\begin{array}{cccc}
     a_1 & a_2 & \cdots & a_{n}\\
    1 & 1 & \cdots & 1 \\
    0 & 0 & \cdots & 0 \\
    0 & 0 & \cdots & 0 \\
\end{array}\right].
$$

\begin{lemma}
    \label{lem:combination of two}
    Let $H_1=[a_1^{1},\cdots,a_{n_1}^{1}]$ be the incidence matrix of an $(m,r-1,n_1)$ DCP-CWGC $C_1$, $H_2=[a_1^{2},\cdots,a_{n_2}^{2}]$ be the incidence matrix of an $(m,r,n_2)$ DCP-CWGC $C_2$ . Suppose $a_{n_1}^{1} \vee a_1^{2} = a_1^{2}$, and $a_{n_1}^{1} \vee a_1^{2} \neq a_j^{1} \vee a_{j+1}^{1}$ for all $1 \leq j \leq n_1-1$, then
    \begin{equation}
    \label{eq:combination of 2}
    H:=[H_1^{+},H_2^{-}]=\left[\begin{array}{cccccc}
    a_1^{1} & \ldots & a_{n_1}^{1} & a_1^{2} & \ldots & a_{n_2}^{2} \\
    1 & \ldots & 1 & 0 &\ldots & 0
    \end{array}\right]        
    \end{equation}
    is the incidence matrix of an $(m+1,r,n_1+n_2)$ DCP-CWGC $C$, which is obtained by the combination of $C_1^{+}$ and $C_2^{-}$.
\end{lemma}

\begin{IEEEproof}
The columns of $H$ are binary vectors of length $m+1$ and weight $r$, and consecutive columns have Hamming distance $2$. Observing the last index of the columns in $H$, it follows that all columns of $H$ and all OR-sums of any two consecutive columns of $H$ are pairwise different. In particular, because of the technical conditions posed on $a_{n_1}^{1} \vee a_1^{2}$ in the premise, $[a_{n_1}^{1} \vee a_1^{2},1]^T$ is distinct from other OR-sums of consecutive columns in $H$.
\end{IEEEproof}

We term $C_1$ and $C_2$ in \cref{lem:combination of two} the elementary DCP-CWGCs for the combination, and their augmented codes, $C_1^{+}$ and $C_2^{-}$, serve as component codes. Utilizing the augmented structure, it is straightforward to generalize \cref{lem:combination of two} to the combination of multiple DCP-CWGCs.

\begin{corollary}
\label{cor:combination of L}
   Let  $\mathfrak{C}=\{C_{m_0},C_{m_0+1},\cdots,C_{m}\}$ be a collection of $L$ elementary DCP-CWGCs, $L=m-m_0+1$, where $m_0<m$ is a pre-specified positive integer, $C_j$ is a $(j-1,r-1,n_j)$ DCP-CWGC, $m_0+1 \leq j \leq m$; and $C_{m_0}$ is an $(m_0,r,n_{m_0})$ DCP-CWGC. Suppose that the technical conditions for the combination in \cref{lem:combination of two} are satisfied, then repeated application of \cref{lem:combination of two} gives the incidence matrix of an $(m,r,\sum_{j=m_0}^{m} n_j)$ DCP-CWGC $C$.
\end{corollary}

\begin{example}
\label{ex:recursive combination}
Let $C_1$ be a simple $(5,1,5)$ DCP-CWGC with the following incidence matrix
    $$H_{5,1,5}=\left[\begin{array}{ccccc}
      1 & 0 & 0 & 0 & 0\\
      0 & 0 & 1 & 0 & 0\\
      0 & 0 & 0 & 0 & 1\\
      0 & 0 & 0 & 1 & 0\\
      0 & 1 & 0 & 0 & 0\\
    \end{array}\right],$$
and $C_2$ be the DCP-CWGC in \cref{ex:incidence_matrices}. $C_1$ and $C_2$ meet the conditions in \cref{lem:combination of two}. Therefore, by combining $C_{1}^{+}$ and $C_{2}^{-}$, we obtain a $(6,2,15)$ DCP-CWGC $C$ with incidence matrix 
    \begin{equation*}
    \begin{aligned}
     &H_{6,2,15}=\\ 
     &\left[\begin{array}{ccccccccccccccc}
      1 & 0 & 0 & 0 & 0 & 0 & 1 & 1 & 0 & 0 & 0 & 0 & 0 & 1 & 1\\
      0 & 0 & 1 & 0 & 0 & 1 & 0 & 0 & 1 & 0 & 0 & 1 & 0 & 0 & 1\\
      0 & 0 & 0 & 0 & 1 & 1 & 1 & 0 & 0 & 1 & 1 & 0 & 0 & 0 & 0\\
      0 & 0 & 0 & 1 & 0 & 0 & 0 & 1 & 1 & 1 & 0 & 0 & 1 & 0 & 0\\
      0 & 1 & 0 & 0 & 0 & 0 & 0 & 0 & 0 & 0 & 1 & 1 & 1 & 1 & 0\\
      1 & 1 & 1 & 1 & 1 & 0 & 0 & 0 & 0 & 0 & 0 & 0 & 0 & 0 & 0\\
    \end{array}\right].    
    \end{aligned}
    \end{equation*}    
\end{example}

Note that the above combination operation in \cref{lem:combination of two} exhibits a recursive nature, where the combined code itself may be used as a component DCP-CWGC for a new combination operation. 
Based on this recursive combination approach, we prove the existence of maximal strictly balanced DCP-CWGCs for all positive integers $r$ and $m \geq r+1$. 
The proof utilizes a useful observation that the incidence matrices of codes in $\text{DCP-CWGC}(m,r,n)$ are invariant under row permutations. \\

\begin{lemma}
    \label{lem:permute}
    Let $H$ be an incidence matrix of an $(m,r,n)$ DCP-CWGC, then for any $m \times m$ permutation matrix $P$, $P \cdot H$ is also an incidence matrix of an $(m,r,n)$ DCP-CWGC.
\end{lemma}

Moreover, given $r$ fixed, one of the starting codes for the combination is chosen as a maximal strictly balanced $\left(2r+1,r,\binom{2r+1}{r}\right)$ DCP-CWGC. Such a code is proven to exist from the well-known \emph{middle two levels problem} \cite{mutze_proof_2014,mutze_book_2024}. It is straightforward to verify that the solution to the middle two levels problem satisfies the constraints in \cref{def:dcp-cwgc}. Therefore, any solution to the middle two levels problems $C$ is a $\left(2r+1,r,n=\binom{2r+1}{r}\right)$ DCP-CWGC and we get the following lemma.\\

\begin{lemma}
    \label{lem:middle two levels}
    There exists a $\left(2r+1,r,n=\binom{2r+1}{r}\right)$ DCP-CWGC for any positive integer $r$.\\
\end{lemma}

\noindent Following on from the previous lemma, we first establish the existence of DCP-CWCGs for any positive integers $r$ and $m\geq 2r+1$.\\

\begin{proposition}\label{prop:maximal_DCP-CWGC}
    There exists an $(m,r,n=\binom{m}{r})$ DCP-CWGC for any positive integer $r$ and $m \geq 2r+1$.\\
\end{proposition}
 
\begin{IEEEproof}
    The proof is by induction on $r$. For $r=1$, it is straightforward that there exists an $(m,1,m)$ DCP-CWGC for any $m \geq 3$. 
    
    For any fixed integer $r \geq 2$, suppose that there exist $\left(m,r-1,\binom{m}{r-1}\right)$ DCP-CWGCs for all $m \geq 2(r-1)+1$, and we denote their incidence matrices as $H_{m,r-1,\binom{m}{r-1}}=[a_1^{(m,r-1)},\cdots,a_{\binom{m}{r-1}}^{(m,r-1)}]$, respectively. Moreover, assume that for all $m \geq 2r+1$, there exists a column vector $y_{m,r-1}$ for $H_{m,r-1,\binom{m}{r-1}}$ such that $y_{m,r-1} \vee a_{\binom{m}{r-1}}^{(m,r-1)}=y_{m,r-1}$ and $y_{m,r-1} \neq a_l^{(m,r-1)} \vee a_{l+1}^{(m,r-1)}$, $\forall 1 \leq l \leq \binom{m}{r-1}-1$, which enables the application of~\cref{cor:combination of L}. 
    
    We will construct an $\left(m,r,\binom{m}{r}\right)$ DCP-CWGC for any $m \geq 2r+1$. The construction is by induction on $m$. By \cref{lem:middle two levels}, there exists a $\left(2r+1,r,\binom{2r+1}{r}\right)$ DCP-CWGC, and we denote its incidence matrix as $H_{2r+1,r,\binom{2r+1}{r}}=[a_1^{(2r+1,r)},\cdots,a_{\binom{2r+1}{r}}^{(2r+1,r)}]$. Since the construction in \cref{lem:middle two levels} comes from the middle two levels problem, there exists a column vector $y_{2r+1,r}$ such that $y_{2r+1,r} \vee a_{\binom{2r+1}{r}}^{(2r+1,r)}=y_{2r+1,r}$ and $y_{2r+1,r} \neq a_l^{(2r+1,r)} \vee a_{l+1}^{(2r+1,r)}$, for all $1 \leq l \leq \binom{2r+1}{r}-1$. 
    
    Suppose that we have constructed an $(m,r,\binom{m}{r})$ DCP-CWGC for some $m \geq 2r+1$, whose $m \times \binom{m}{r}$ incidence matrix $H_{m,r,\binom{m}{r}}=[a_1^{(m,r)},a_2^{(m,r)},\cdots,a_{\binom{m}{r}}^{(m,r)}]$ contains a column vector $y_{m,r}$ satisfying $y_{m,r} \vee a_{\binom{m}{r}}^{(m,r)}=y_{m,r}$ and $y_{m,r} \neq a_l^{(m,r)} \vee a_{l+1}^{(m,r)}$, for all $1 \leq l \leq \binom{m}{r}-1$. 
    From the induction hypothesis on $r$, there exists an $\left(m,r-1,\binom{m}{r-1}\right)$ DCP-CWGC, whose incidence matrix is denoted as $H_{m,r-1,\binom{m}{r-1}}$, and contains a corresponding column vector $y_{m,r-1}$. Thus, we may apply a suitable row permutation $P$ on $H_{m,r,\binom{m}{r}}$, giving us $H_{m,r,\binom{m}{r}}^{'}=P\cdot H_{m,r,\binom{m}{r}}$, such that $P \cdot a_1^{(m,r)} = y_{m,r-1}$. Then, by \cref{lem:combination of two}, 
    \begin{equation*}
        \begin{aligned}
            H_{m+1,r,\binom{m+1}{r}}&=[a_1^{(m+1,r)}\cdots,a_{\binom{m+1}{r}}^{(m+1,r)}]\\
    &:=\left[\begin{array}{cc}
    H_{m,r-1,\binom{m}{r-1}} & H_{m,r,\binom{m}{r}}^{'}\\
    \mathbf{1} & \mathbf{0}\end{array}\right]
        \end{aligned}
    \end{equation*}
    gives the incidence matrix of an $\left(m+1,r,\binom{m+1}{r}\right)$ DCP-CWGC. Moreover, the column vector $y_{m+1,r}= (P\cdot y_{m,r})^{-}$ satisfies $y_{m+1,r} \vee a_{\binom{m+1}{r}}^{(m+1,r)}=y_{m+1,r}$ and $y_{m+1,r} \neq a_l^{(m+1,r)} \vee a_{l+1}^{(m+1,r)}$, for all $1 \leq l \leq \binom{m+1}{r}-1$. 
    
    Thus, for all $m \geq 2r+1$, we have constructed an $\left(m,r,\binom{m}{r}\right)$ DCP-CWGC, and the proof is completed. 
\end{IEEEproof}

\vspace{10pt}

Then, we utilize an observation that a path between $S(m,r)$ and $S(m,r+1)$ is isomorphic to a path between $S(m,m-r)$ and $S(m,m-r-1)$ to prove that DCP-CWCGs exist for any positive integers $r$ and $m\geq r+1$.
 
\begin{theorem}\label{thm:maximal_DCP-CWGC}
    There exists an $(m,r,n=\min\{\binom{m}{r},\binom{m}{r+1}+1\})$ DCP-CWGC that meets the theoretical bound in \cref{prop:bound} for any positive integers $r$ and $m \geq r+1$.\\
\end{theorem}
\begin{IEEEproof}
    From \cref{prop:maximal_DCP-CWGC}, for any $m \geq 2r+1$, there exists an $\left(m,r,\binom{m}{r}\right)$ DCP-CWGC $C$ whose incidence matrix $H_{m,r,\binom{m}{r}}=[a_1^{(m,r)},\cdots,a_{\binom{m}{r}}^{(m,r)}]$. Moreover, there exists a column vector $y_{m,r}$ that satisfies $y_{m,r} \vee a_{\binom{m}{r}}^{(m,r)}=y_{m,r}$ and $y_{m,r} \neq a_l^{(m,r)} \vee a_{l+1}^{(m,r)}$, for all $1 \leq l \leq \binom{m}{r}-1$. Thus, we may establish a path of length $2 \cdot \binom{m}{r} +1$ between $S(m,r)$ and $S(m,r+1)$, which contains $\binom{m}{r}$ nodes in $S(m,r)$ and $\binom{m}{r} +1$ nodes in $S(m,r+1)$. By flipping all the ``1''s to ``0''s and ``0''s to ``1''s, it gives a path of length $2 \cdot \binom{m}{r} +1$ between $S(m,m-r)$ and $S(m,m-r-1)$, which contains $\binom{m}{m-r}$ nodes in $S(m,m-r)$ and $\binom{m}{m-r} +1$ nodes in $S(m,m-r-1)$. This path is identical to a $(m,m-r-1,n=\binom{m}{m-r} +1)$ DCP-CWGC $\bar{C}$. Since $2(m-r-1)+1 \leq 2r+1 \leq m$, $\bar{C}$ also meets the theoretical bound in \cref{prop:bound} and thus the theorem holds.
\end{IEEEproof}

\subsection{Recursive Combination together with BBA}

\begin{figure*}[htbp]
\centering
\includegraphics[width=0.8\textwidth]{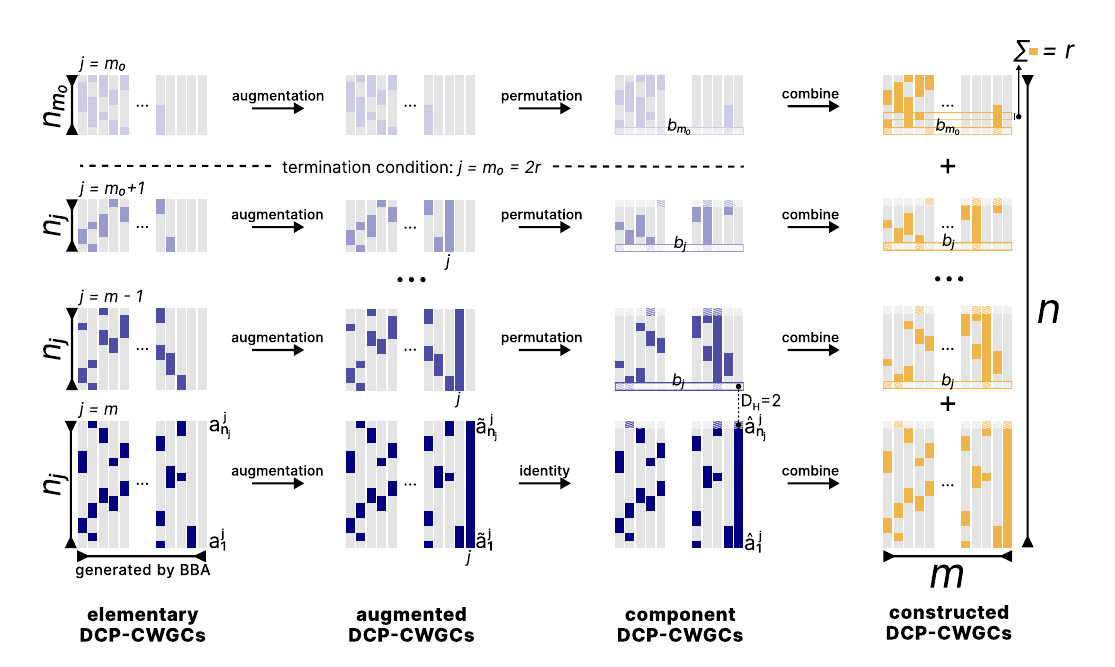}
\caption{{\bfseries The rcBBA scheme.} rcBBA is based on iterative construction of the DCP-CWGC with parameters $(m, r, n)$ from elementary DCP-CWGCs. rcBBA iterates with iteration counter $j$, which starts at $j = m$ and decrements by 1 with each iteration. Each iteration (row) has three main steps (columns): 1) generation of the elementary DCP-CWGC with BBA with parameters $(j-1, r-1, n_j)$, 2) augmentation, and 3) permutation such that the Hamming distance between the first address of the generated DCP-CWGC component $b_j$ and the last address of base code is equal to 2. After all three steps, a component is added to the base code.  When rcBBA reaches $j = 2r$, it switches to the final iteration regime, where it generates elementary DCP-CCWGC with parameters $(m_0, r, n_{m_0})$, augments it, permutes, and adds to the base code. }
\label{fig:rcbba}
\end{figure*}

We paired the recursive combination strategy with BBA, which we will henceforth call rcBBA: BBA generates elementary DCP-CWGCs, which are then augmented and combined using the recursive combination. This way, rcBBA can construct balanced DCP-CWGCs with flexible parameters $(m,r,n)$.

rcBBA is based on the iterative construction of the full $(m,r,n)$ DCP-CWGC. Each iteration consists of three main steps. 
\begin{enumerate}
    \item[1] The length of the elementary DCP-CWGC is determined.
    \item[2] The elementary DCP-CWGC is generated by BBA (\cref{fig:rcbba}, ``elementary DCP-CWGCs'' column) and then transformed into a DCP-CWGC component by augmentation (\cref{fig:rcbba}, ``augmented DCP-CWGCs'' column) and permutation (\cref{fig:rcbba}, ``component DCP-CWGCs'' column).
    \item[3] The DCP-CWGC component is combined into the base code (\cref{fig:rcbba}, ``constructed DCP-CWGCs'' column).
\end{enumerate}

The next section provides a detailed description of the rcBBA core function \texttt{RecCombine} (\cref{fun:reccom}), which is applied in each iteration.

First, we will introduce the notation used in this section. $j$ serves as an iteration counter, rcBBA starts with $j = m$, then $j$ is decreased by $1$ with each iteration, and when rcBBA reaches $\binom{j-1}{r} \leq \binom{j-1}{r-1}$, i.e., $j \leq 2r$, it switches to the final iteration regime.
Thus, when $m \leq m_0=2r$, rcBBA jumps directly into the final iteration regime and is completed with BBA.
Reasons for selecting $m_0=2r$ as the termination condition are discussed in \cref{sec:results}. At the beginning of each iteration, a new residual balance vector $W_{res}$ is determined. The length of the elementary DCP-CWGC at the $j$-th iteration is denoted as $n_j$. The elementary DCP-CWGC is denoted as $C_j$ and the corresponding incidence matrix is denoted as $H_j$. Addresses in $C_j$ are denoted as $a^{j}_{i}$, where $i = {1, 2, ..., n_j}$. The $k$-th element of the address $a^{j}_{i}$ is denoted as $a^{j}_{i, k}$. We denote the base incidence matrix as $H$ with the $i$-th address $a^{H}_{i}$, where $i = {1, 2, ..., n_H}$. We denote an augmentation procedure and permutation operation by adding `` $\tilde{}$ '' and `` $\hat{}$ '' respectively: for example, the augmented DCP-CWGC  $\tilde{C_j}$ and the permuted DCP-CWGC $\hat{C_j}$. 

\begin{algorithm}[hbt!]
\renewcommand{\thealgorithm}{2}
\caption{Recursive Combination together with BBA}\label{alg:rcBBA}
\begin{algorithmic}[1]
\Require the desired parameters $(m,r,n)$ of DCP-CWGC.
\Ensure the incidence matrix of desired DCP-CWGC $H$.
\State initialize iteration counter $j = m$.
\State initialize the residual balance vector $W_{res}= (w_1, w_2, \ldots, w_m)$ by \cref{eq:original_balance}.
\State initialize the residual index set $I_{res} = \{1, 2, ..., j\}$.
\State let $n_j = w_j$.
\State initialize the base incidence matrix $H$ as an empty matrix.
\State $H=\texttt{RecCombine}(H, j, W_{res}, I_{res}, n_j, m, r, n)$.
\State \Return $H$.
\end{algorithmic}
\end{algorithm}

rcBBA with parameters $(m,r,n)$ (\cref{alg:rcBBA}) starts with the initialization of the iteration counter $j$: $j = m$. Then residual balance vector $W_{res}$ is initialized: $W_{res} = (w_1, w_2, \ldots, w_m)$, where $w_i$ is calculated using the floored average value $w = \lfloor \frac{r \cdot n}{m} \rfloor$. If $\frac{r \cdot n}{m}$ is not an integer, then 1 is added to the first $r \cdot n - m \cdot w$ elements:
\begin{equation}
\label{eq:original_balance}
    w_i = \begin{cases}
    w+1, \qquad  & 1 \leq i \leq r \cdot n - m \cdot w\\
    w, \qquad & r \cdot n - m \cdot w < i \leq m
\end{cases}.
\end{equation}
Also, the residual index set $I_{res}$ is initialized: $I_{res} = \{1, 2, ..., j\}$.

Next, the length of the elementary DCP-CWGC for $j$-th iteration $n_j$ is determined as $j$-th element of $W_{res}$: $n_j = w_j$.
The parameters for the next iteration ($H, j, W_{res}, I_{res}, n_j, m, r, n$) are then entered into the \texttt{RecCombine} function ( \cref{fun:reccom}).

\begin{function}[hbt!]
\caption{\texttt{RecCombine}($H, j, W_{res}, I_{res}, b_j, n_j, m, r, n$)}
\label[function]{fun:reccom}
\begin{algorithmic}[1]
\Require base incidence matrix $H$, iteration counter $j$, residual balance vector $W_{res}$, residual index set $I_{res}$, the first binary address for the next DCP-CWGC $b_j$ (optional), length of the next DCP-CWGC $n_j$; number of pools $m$, weight of binary address $r$, and desired code length $n$.
\Ensure the incidence matrix of the constructed DCP-CWGC $H$.
\State termination condition: $m_0 =  2r$.
\If{$j = m_0$}
    \State $n_{m_0} = n - n_H$.
    \State randomly select $a^{m_0}_{1} \in S(m_0, r)$
    \State find a permutation map $P$ on $\{1,\cdots,m\}$ such that it satisfies (\ref{eq:constraint m0}).
    \State permute $W_{res}$ with $P^{-1}$: $W_{des} = P^{-1}(W_{res})$
    \State $C_{m_0}=\texttt{BBA}(m_0, r, n_{m_0}, a^{m_0}_{1}, W_{des})$. Its incidence matrix: $H_{m_0}$.
    \If{$H_{m_0} \neq \phi$}
        \State augmentation: $\tilde{H}_{m_0} = H_{m_0}^{(m-m_0)-}$.
        \State row permutation: $\hat{H}_{m_0} = P(\tilde{H}_{m_0})$.
        \State \Return $[H,\hat{H}_{m_0}]$.
    \EndIf
\ElsIf{$j > m_0$}
    \If{$n_H = n$}
        \State return $H$.
    \EndIf
    \State $C_j=\texttt{BBA}(j-1, r-1, n_j)$. Its incidence matrix: $H_j$.
    \If{$H_j \neq \phi$}
        \State augmentation: $\tilde{H}_j = H_{j}^{+,(m-j)-}$.
        \State find a permutation map $P$ on $\{1, \cdots, m\}$, such that (\ref{eq:constraint j}) is satisfied.
        \State row permutation: $\hat{H}_j=P (\tilde{H}_j)$.
        \State update base incidence matrix: $H = [H,\hat{H}_j]$.
        \State update $W_{res}$: $W_{res} = W_{res} - W_{H_j}$.
        \State update $j$: $j = j - 1$
        \State search the set $B_j$ consisting of binary vectors $b_j$'s that satisfy (\ref{eq:condition_of_b}).
        \If{$B_{j} \neq \phi$}
            \State for each $b_j$, find corresponding $n_j = W_{res}[I(b_j)[r]]$ and sort $b_j$ according to their $n_j$ in the descending order.
            \For{$b_j$ in sorted $B_j$}
                \State $H=\texttt{RecCombine}(H, j, W_{res}, I_{res}, b_j,$ $ n_j, m, r, n)$.
    \State \Return $H$.
    \EndFor
    \EndIf
\EndIf
\EndIf
\end{algorithmic}
\end{function}

In each iteration of \texttt{RecCombine} except for the first one, a new DCP-CWGC component is generated. We must ensure the proper joining of this new component with the existing base incidence matrix so that it will satisfy constraints \cref{item:constraint1}, \cref{item:constraint2}, and \cref{item:constraint3}. Consequently, we need to search for a suitable binary vector $b_j$ which is later used to find a permutation map to turn an augmented DCP-CWGC into a DCP-CWGC component. In this procedure, $b_j$ will become the first address of the new component DCP-CWGC $\hat{C_j}$, i.e., the first column of its incidence matrix $\hat{H_j}$. In the first iteration, $b_j$ is not used.

Suitable $b_j$ should satisfy the following requirements~\refstepcounter{equation}(\theequation)\label{eq:condition_of_b}:
\begin{itemize}
    \item $b_j$ is a binary vector with length $m$ and weight $r$:
    $$b_j \in S(m, r)$$
    \item each element in the index form of $b_j$ is in the residual index set $I_{res}$: $$\forall i \in I(b_j): i \in I_{res}$$
    \item The Hamming distance between $b_j$ and the last address in $H$ $a_{n_H}^{H}$ equals to $2$, i.e. $b_j \in Adj(Adj(a^{H}_{n_H}))$:
    $$D_H(b_j, a^{H}_{n_H})=2$$
    \item the union formed by $a^{H}_{n_H}$ and $b_j$ is distinct from consecutive unions in $H$:
    $$\forall k = {1, 2, ..., n_{H}-1}: a^{H}_{n_H} \vee b_j \neq a^{H}_{k} \vee a^{H}_{k+1}$$
\end{itemize}
We denote the set of vectors $b_j$ satisfying these requirements as $B_j$.

\texttt{RecCombine} has two regimes: (1) when $j > m_0$; and (2) when $j = m_0$. If \texttt{RecCombine} reaches $j = m_0$, it stops iterating through $j$ and switches to the final iteration regime.

\subsubsection{if $j>m_0$}
    \begin{itemize}
        \item generate a $(j-1,r-1,n_j)$ elementary DCP-CWGC $C_j$ by BBA with the incidence matrix $H_j$:
        $$H_j=[a^{j}_{1}, \cdots, a^{j}_{n_j}].$$
        \item apply augmentation on $H_j$: $$\tilde{H}_j=[\tilde{a}^{j}_{1}, \cdots, \tilde{a}^{j}_{n_j}] := H_{j}^{+,(m-j)-}.$$
        \item find a permutation map $P$ on $\{1, 2, \cdots, m\}$, such that~\refstepcounter{equation}(\theequation)\label{eq:constraint j}:
        \begin{itemize}
            \item if it is the first iteration ($j = m$), the permutation map is an identity matrix;
            \item $P$ maps elements in $\{1,\cdots,j\}$ one-to-one into $I_{res}$: $$P(\{1,\cdots,j\})= I_{res};$$
            \item the first address of $\tilde{H}_j$ is permuted to $b_j$: $$\tilde{a}^j_{1} \to b_j;$$
            \item $r$-th element of the $I(\tilde{a}^j_{1})$ is permuted to the $r$-th element of $I(b_j)$: $$I(\tilde{a}^j_{1})[r] \to I(b_j)[r].$$
        \end{itemize}
        \item apply $P$ as a row permutation on $\tilde{H}_j$: 
        $$\hat{H}_j=[\hat{a}^{j}_{1},\cdots,\hat{a}^{j}_{n_j}]=P (\tilde{H}_j).$$
    \item add $\hat{H}_j$ to the base incidence matrix: $$H \gets [H, \hat{H}_j]$$
    \item update parameters to prepare for the next iteration: remove the $r$-th element of $I(b_j)$ from the residual index set $I_{res}$: $I_{res} \gets I_{res} \backslash \{I(b_j)[r]\}$, update residual balance vector $W_{res}$ by subtracting the balance vector of the new incidence matrix: $W_{res} \gets W_{res} - W_{\hat{H}_j}$, update 
    $j = j-1$;
    \item find suitable $b_j \in B_j$ that satisfies \cref{eq:condition_of_b};
    \item for each $b_j$, find corresponding $n_j$ equal to $I(b_j)[r]$-th element of $W_{res}$, where $I(b_j)[r]$ is the $r$-th element of the index form of $b_j$: $$n_j = W_{res}[I(b_j)[r]].$$ Sort $B_j$ based on the $n_j$ in descending order.
    \item for each $b_j$ in sorted $B_j$, continue with new parameters: $H, j, W_{res}, I_{res}, b_{j}, n_j$.
    \end{itemize}

\subsubsection{$j=m_0$}
    \begin{itemize}
        \item determine the length of the last component DCP-CWGC $C_{m_0}$ with incidence matrix $H_{m_0}$ by subtracting number of columns of $H$ from the desired length $n$: $$n_{m_0} = n - n_H;$$
        \item randomly select a first address $a^{m_0}_{1} \in S(m_0, r);$
        \item find a permutation map $P$ on $\{1,\cdots,m\}$, such that~\refstepcounter{equation}(\theequation)\label{eq:constraint m0}:
            \begin{itemize}
                \item if $j = m$, permutation map is an identity matrix;
                \item $\{1,\cdots,m_0\}$ is mapped one-to-one to $I_{res}$: $$P(\{1,\cdots,m_0\})=I_{res};$$
                \item the first address $a^{m_0}_{1}$ is permuted to $b_j$: $$a^{m_0}_{1} \to b_j.$$
            \end{itemize}
        \item determine desired balance vector $W_{des}$ by applying inverse index permutation map $P^{-1}$ on the residual balance vector $W_{res}$: $$W_{des} = P^{-1}(W_{res}).$$
        \item generate by BBA an $(m_0, r, n_{m_0})$ elementary DCP-CWGC $C_{m_0}$ with desired balance $W_{des}$, whose incidence matrix $H_{m_0}$:
        $$H_{m_0} =[a^{m_0}_{1}, \cdots, a^{m_0}_{n_{m_0}}].$$
        \item apply augmentation on $H_{m_0}$: $$\tilde{H}_{m_{0}}=[\tilde{a}^{m_0}_{1},\cdots, \tilde{a}^{m_0}_{n_{m_{0}}}]:=H_{m_{0}}^{(m-m_{0})-}.$$
        \item apply $P$ as a row permutation on $\hat{H}_{m_0}$: $$\hat{H}_{m_0}=[\hat{a}^{m_0}_{1},\cdots,\hat{a}^{m_0}_{n_{m_0}}] = P (\tilde{H}_{m_0}).$$
        \item update the base incidence matrix $H \gets [H,\hat{H}_{m_0}]$, and return $H$ as the incidence matrix of the constructed DCP-CWGC $C$.
    \end{itemize}

Note that there may be no binary vector $b_j$ satisfying \cref{eq:condition_of_b} in some iteration $j=m-1,\cdots,m_0$, thus the above iterative process may reach a dead end. At this point, the algorithm needs to backtrack to the previous iteration, choose other $b_j \in B_j$, and using that $b_j$ find another component DCP-CWGC to replace the current DCP-CWGC.  

Due to the augmented structure of component DCP-CWGCs and that their lengths are recursively determined in each iteration, we have the following identities:
\begin{equation}
  \label{eq:lengths}
  \begin{cases}
    n_{j} = W_{ini}[I(b_j)[r]] - \sum_{j'=j+1}^{m} W_{\hat{H}_{j'}}[I(b_j)[r]],\\
    \qquad \qquad \qquad \qquad \qquad \quad \forall j=m,\cdots,m_0+1;\\
    n_{m_0}=n-\sum_{j=m_0+1}^n n_j;
  \end{cases}
\end{equation}
where $W_{ini}$ is the initialized residual balance vector. Note that $W_{ini}$ has deviation at most $1$, and the lengths of the component DCP-CWGCs are determined based on the current residual balance vector in each iteration. In sight of this, we can prove that the combined DCP-CWGC is nearly balanced if $H_{m_0}$ has a balance vector that approximates $W_{des}$ in the $m_0$-th iteration:

\begin{theorem}
  The deviation of the combined DCP-CWGC $C$ is upper bounded by 
  $$\delta_{C} \leq 2 \cdot \max_{i=1,\cdots,m_0} \{|W_{H_{m_0}}[i]-W_{des}[i]|\}+2.$$
\end{theorem}

\begin{IEEEproof}
  Since $W_{ini}$ is optimally initialized, and that the lengths of component DCP-CWGCs are determined by \cref{eq:lengths}, it is straightforward to show that $W_{H}(I(b_m)[r]), \cdots,W_{H}(I(b_{m_0+1})[r]) \in \{w,w+1\}$, with $w=\lfloor \frac{r \cdot n}{m} \rfloor$. Denote 
  \begin{equation*}
    \begin{aligned}
        I_{res}^{(m_0)} &=\{i_1,\cdots,i_{m_0}\} \\
        &= \{1,\cdots,m\} \backslash \{I(b_{m})[r],\cdots,I(b_{m_0+1})[r]\}.
    \end{aligned}      
  \end{equation*}
  Moreover, we denote $W_{res}$ to be the residual balance vector in the $m_0$-th iteration, and denote $P$ to be the permutation map in the $m_0$-th iteration.
  Then, we have 
  \begin{align*}
    \delta_{C} &= \max_{i = 1,\cdots,m} \{W_{H}[i]\} - \min_{i =1,\cdots,m} \{W_{H}[i]\}\\
    & \leq \max_{i \in I_{res}^{(m_0)}} \{W_{H}[i]\} - \min_{i \in I_{res}^{(m_0)}} \{W_{H}[i]\} +1\\
    & = \max_{i \in I_{res}^{(m_0)}} \left\{W_{\hat{H}_{m_0}}[i] + \sum_{j=m_0+1}^{m} W_{\hat{H}_{j}}[i]\right\} \\
    & \qquad \qquad - \min_{i \in I_{res}^{(m_0)}} \left\{W_{\hat{H}_{m_0}}[i] + \sum_{j=m_0+1}^{m} W_{\hat{H}_{j}}[i]\right\} +1 \\
    & = \max_{i \in I_{res}^{(m_0)}} \left\{W_{\hat{H}_{m_0}}[i] + W_{ini}[i] - W_{res}[i]\right\} \\
    & \qquad \qquad - \min_{i \in I_{res}^{(m_0)}} \left\{W_{\hat{H}_{m_0}}[i] + W_{ini}[i] - W_{res}[i]\right\} +1 \\
    & \leq \max_{i \in I_{res}^{(m_0)}} \left\{W_{\hat{H}_{m_0}}[i] - W_{res}[i]\right\} + \max_{i \in I_{res}^{(m_0)}} \{W_{ini}[i]\}\\
    & - \min_{i \in I_{res}^{(m_0)}} \left\{W_{\hat{H}_{m_0}}[i] - W_{res}[i]\right\} - \min_{i \in I_{res}^{(m_0)}} \{W_{ini}[i]\} +1 \\
    & \leq 2 \cdot \max_{i \in I_{res}^{(m_0)}} \left\{|W_{\hat{H}_{m_0}}[i]-W_{res}[i]|\right\} +2\\
    & = 2 \cdot \max_{i \in P^{-1}(I_{res}^{(m_0)})} \left\{|W_{H_{m_0}}[i]-W_{des}[i]|\right\}+2\\
    & = 2 \cdot \max_{i=1,\cdots,m_0} \left\{|W_{H_{m_0}}[i]-W_{des}[i]|\right\}+2.
  \end{align*}
\end{IEEEproof}

\section{Computational Complexity and Simulation Results}
\label{sec:results}

Both BBA and rcBBA are implemented in our open-source software \textit{codePub} (detailed documentation at \url{https://codepub.readthedocs.io/}). We evaluated their performance in terms of computational complexity and run time, balance of the DCP-CWGCs, and error detection ability.

\subsection{Computational Complexity and Empirical Run Time}

The computational complexity of BBA is mainly driven by the heuristic balance-optimized path search. Consider a balanced $(m,r,n)$ DCP-CWGC constructed by BBA. For simplicity, we assume that the computational complexity when visiting each node (including computing the variances, sorting the variance vector, etc.) is the same ``unit'' value.

Under this assumption, the overall computational complexity of BBA is directly proportional to the number of nodes that are visited during the path search. Thus, the best-case computational complexity of BBA is linear in $n$ ($\Theta(n)$).
In the worst case, the path is not updated until the last candidate node is checked. Each address $a$ has at most $m-r$ choices in $Adj(a)$ for the next union $u$, and each union $u$ has at most $r$ choices in $Adj(u)$ for the next address $a$. Thus, the worst-case time complexity is upper bounded by $O([r(m-r)]^n)$.
In rcBBA, the lengths of elementary DCP-CWGCs are much smaller than the desired full length and thus, it has significantly lower worst-case complexity than full-length construction for BBA. Moreover, the auxiliary overhead associated with the recursive combination procedure (such as augmentation, permutation and combination) is relatively small compared to the complexity on generating elementary DCP-CWGCs by BBA. Therefore, rcBBA generally exhibits lower overall computational complexity than BBA. 

\begin{figure}[htbp]
\centering
\includegraphics[width=0.35\textwidth]{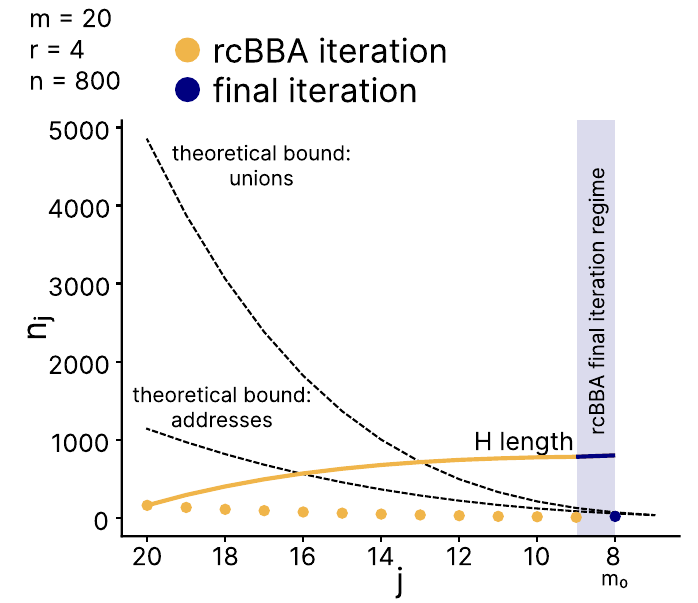}
\caption{{\bfseries The length of the DCP-CWGC component $n_j$ generated with each iteration $j$.} Due to rcBBA's iterative nature with decreasing $j$, the theoretical bound for the length of the component codes, i.e., $\text{min}\{\binom{j}{r-1}, \binom{j}{r}+1\}$, also decreases with each iteration. To avoid reaching this theoretical bound before construction is completed, rcBBA switches to the final iteration regime at an arbitrary $j = 2r$. Orange dots represent elementary codes with length $n_j$ produced during each rcBBA iterations, the blue dot represents the elementary code with length $n_j$ produced at the last iteration. The solid line is the cumulative length of the arrangement.}
\label{fig:m0_selection}
\end{figure}

Of note, with each iteration of rcBBA, the value of $j$ decreases, leading to the construction of elementary DCP-CWGCs with progressively smaller $m$. As a result, rcBBA may eventually reach an iteration where the length of the next elementary sequence, $n_j$, exceeds the maximum possible number of addresses for given $j$ and $r-1$, specifically when $n_j > \binom{j}{r}$. As this point is approached, the probability of rcBBA failing becomes increasingly high. To prevent this outcome, we included the following stopping criterion: when $\binom{j-1}{r} \leq \binom{j-1}{r-1}$ (\cref{fig:m0_selection}), i.e., $j=2r$, rcBBA switches to the final iteration regime.

\begin{figure}[htbp]
\centering
\includegraphics[width=0.35\textwidth]{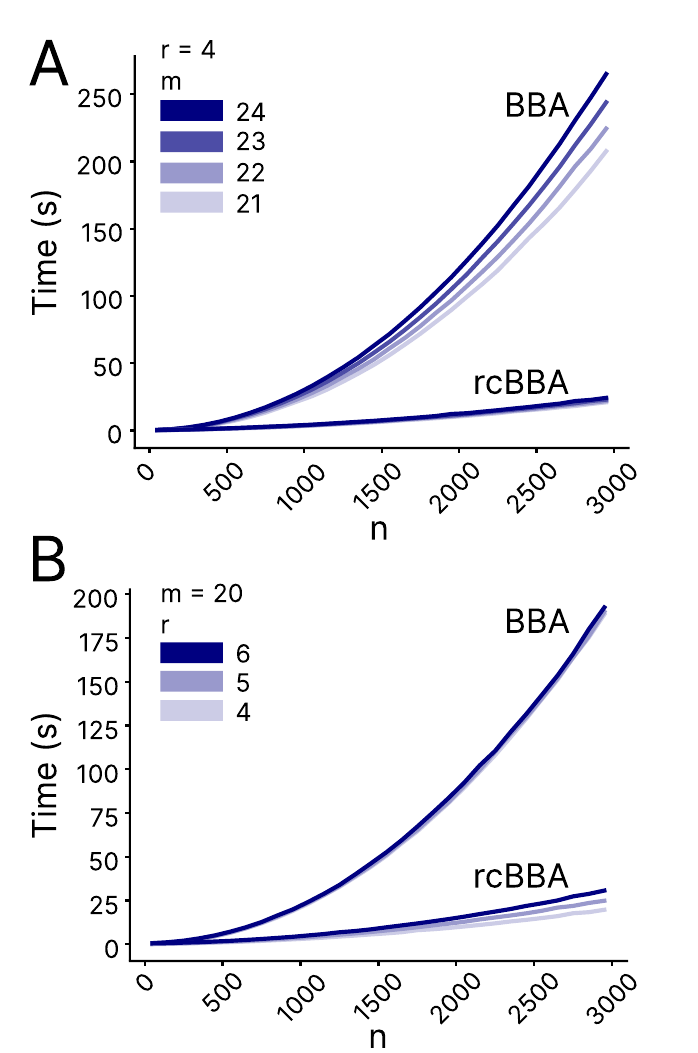}
\caption{\textbf{Empirical runtime analyses for BBA and rcBBA.} Runtime is evaluated as a function of code length $n$, with varying address lengths $m$ (A) and address weights $r$ (B). The process was executed on a single core of a general-purpose CPU node equipped with dual Xeon 6252 processors. It utilized 16.5 GB of virtual memory, with 69.4 MB resident in RAM.}
\label{fig:runtime}
\end{figure}

We tested the performance of BBA and rcBBA by recording their run time across a wide range of parameters. Overall, rcBBA demonstrated significantly faster performance than BBA, especially in the regime of large $n$ (\cref{fig:runtime}). The run time for both algorithms was affected by address length $m$ and address weight $r$; as $n$ increased, $m$ had a greater impact on run time than $r$ for BBA, while $r$ has a greater impact on runtime than $m$ for rcBBA (\cref{fig:runtime}A versus B). 

\begin{figure}[htbp]
\centering
\includegraphics[width=0.35\textwidth]{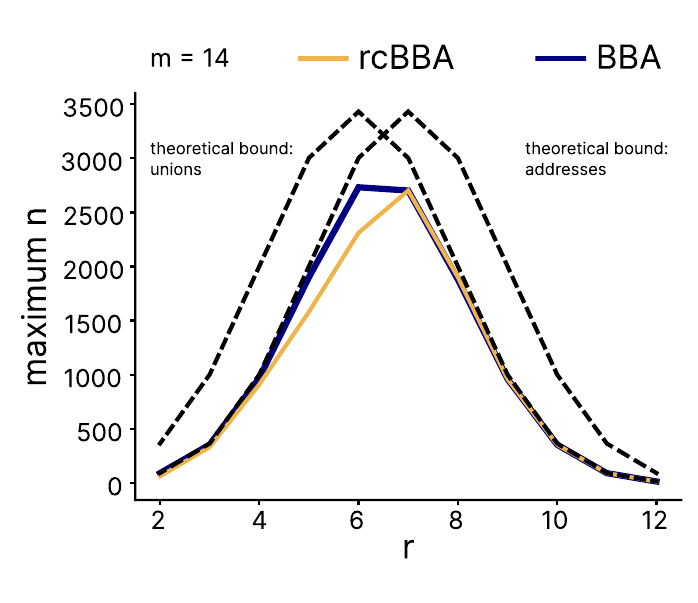}
\caption{\textbf{Performance of BBA and rcBBA near the theoretical bound.} The theoretical bound lies at the combinatorial coefficient for $n$ with  $\text{min}\{\binom{m}{r}, \binom{m}{r+1}+1\}$.  For a fixed number of pools $m$, BBA and rcBBA approach the theoretical bound with increasing address weights $r$.}
\label{fig:theoreticalbound}
\end{figure}

The observed run times for both algorithms were much shorter than those estimated from the worst-case computational complexity. However, as $n$ approached the theoretical bound  $n_{max}=\text{min}\left\{\binom{m}{r}, \binom{m}{r+1}+1\right\}$, the run times increased steeply.
To evaluate performance near the theoretical bound, we tested both algorithms with values of $n$ ranging from $\lfloor{0.6 \cdot n_{max}}\rfloor$ to $\lfloor{0.99 \cdot n_{max}}\rfloor$ , covering $m$ values from 10 to 15 and $r$ values from 2 to 13. If the algorithm returned the code as an empty set or failed to produce a result within the time limit determined by run-time analysis (400 seconds for tested $m$ and $r$ values), we considered such a test as a failure.
For small $r$, BBA slightly outperformed rcBBA, producing a DCP-CWGC with larger $n$ for given $m$ and $r$, although with a much longer runtime; for $r \geq 7$, rcBBA and BBA have the same performance, since rcBBA directly jumps into the final iteration regime which is filled with BBA (\cref{fig:theoreticalbound}).

\subsection{Balance}

\begin{figure}[htbp]
\centering
\includegraphics[width=0.35\textwidth]{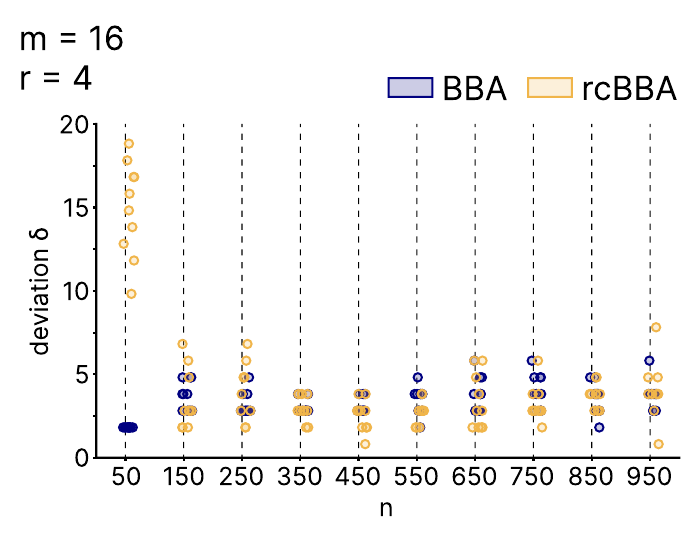}
\caption{\textbf{Empirical balance of pooling schemes with DCP-CWGCs.} Comparison of the deviations $\delta_C = \max(W_C) - \min(W_C)$ of BBA-derived and rcBBA-derived experimental balance. Each dot represents one DCP-CWGC with a deviation $\delta_C$. A total of 10 runs are shown per parameter set ($m, r, n$).}
\label{fig:balance}
\end{figure}

We assessed BBA and rcBBA for their ability to generate balanced DCP-CWGCs by measuring the deviation from the perfect balance for different values of $n$ (\cref{fig:balance}).  For very small $n$ (e.g. $n=50$) BBA vastly outperforms rcBBA in generating balanced arrangements. In contrast, both algorithms produced codes with near optimal balance, with deviations not exceeding $8$ for $n$ ranging from $150$ to $950$.

\subsection{Error Detection}
In a combinatorial pooling experiment with overlapping sequences as items, the number of candidate items is determined by the relative length of the overlap between items and the length of the sequence that generates an experimental signal. If these lengths are identical, then given no experimental errors, the number of target item candidates will be exactly 2. Furthermore, in a combinatorial pooling experiment based on a balanced DCP-CWGC, the number of positive pools remains constant for any positive pair and equals $r+1$ due to constant-weight \cref{item:constraint2} and adjacent distance \cref{item:constraint3} constraints. Consequently, if the number of observed positive pools is not equal to $r+1$, an experimental error must have occurred, with false positive or false negative results if the number of positive pools is greater or less than $r+1$, respectively. This deviation from the expected enables error detection and lets us narrow down the list of potential item candidates based on the positive pools.

We assessed rcBBA's ability to identify two consecutive positive items in the simulation of biological experiments with and without false-negative experimental errors (i.e. erroneously non-identified pools).

In case of an error, the number of potential item candidates depends on:
\begin{itemize}
    \item $n$ (the more items are tested, the more items are in the list of potential candidates);
    \item the number of errors (the more errors the algorithm detected, the fewer pools were activated and consequently the more ambiguous the results are);
    \item the balance of the DCP-CWGC (the more balanced distribution results in a smaller candidate list).
\end{itemize}

\begin{figure}[htbp]
\centering
\includegraphics[width=0.35\textwidth]{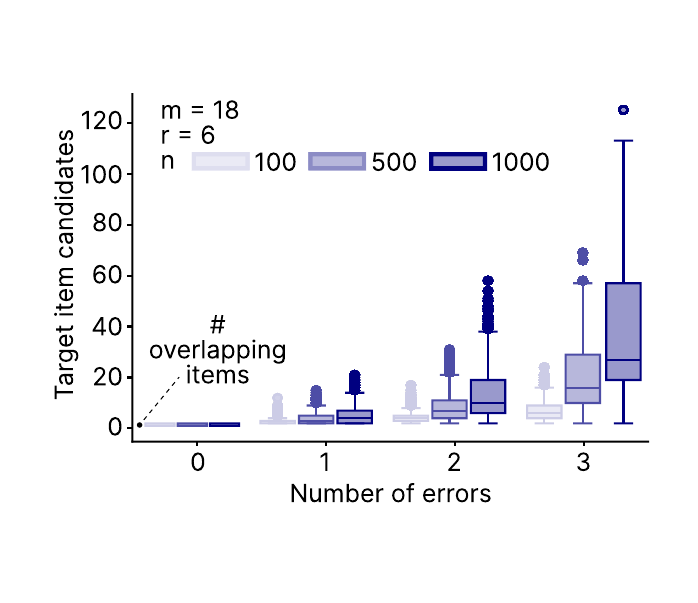}
\caption{\textbf{Experimental error detection with DCP-CWGCs-derived pooling schemes}. For a fixed number of pools $m = 18$ and address weight $r = 6$, the number of target item candidates depends on the number of tested items $n$ and the number of erroneous non-positive pools. Each set of parameters was tested once; for every item, the results of all possible errors were simulated.}
\label{fig:errors}
\end{figure}

We simulated combinatorial pooling experiments using $(18,6,n)$ DCP-CWGCs, where $n$ ranged from 100 to 1000, and tested different levels of false-negative experimental errors. We then quantified the number of candidate items expected in these experiments as a function of the number of errors (\cref{fig:errors}). 
Overall, due to rcBBA's balanced distribution, the candidate list remains small ($\leq 5\%$). For example, with one erroneous non-activated pool, the algorithm can narrow the candidate list to approximately 5 items (for $n$ = 100) or up to 30 items (for $n$ = 1000).

In summary, BBA alone has a better balance for smaller $n$, with a running time similar to rcBBA. However, as $n$ increases, rcBBA achieves a balance consistent with BBA while maintaining faster performance overall.

\section{Conclusion}
\label{sec:conclusion}

We have introduced balanced DCP-CWGCs as a new encoding for combinatorial pooling experiments. DCP-CWGCs can identify any two consecutive positive items, keep the number of tests on each item and each pair of consecutive items constant, and facilitate error detection. 
We proved that there exist maximal strictly balanced $(m,r,n=\binom{m}{r})$ DCP-CWGCs for all positive integer $r$ and $m \geq r+1$. For the efficient construction of balanced DCP-CWGCs with flexible parameters outside this range, we implemented a BBA. Building on BBA, we developed a recursive combination approach, rcBBA, which enables the construction of balanced DCP-CWGCs with larger lengths at great efficiency. Both BBA and rcBBA are implemented in an open-source software \textit{codePub}, and simulation results show that they construct balanced DCP-CWGCs with lengths of hundreds or thousands in tractable time. 

\bibliographystyle{IEEEtran}
\bibliography{bibliography}

\begin{thebibliography}{10}
\providecommand{\url}[1]{#1}
\csname url@samestyle\endcsname
\providecommand{\newblock}{\relax}
\providecommand{\bibinfo}[2]{#2}
\providecommand{\BIBentrySTDinterwordspacing}{\spaceskip=0pt\relax}
\providecommand{\BIBentryALTinterwordstretchfactor}{4}
\providecommand{\BIBentryALTinterwordspacing}{\spaceskip=\fontdimen2\font plus
\BIBentryALTinterwordstretchfactor\fontdimen3\font minus \fontdimen4\font\relax}
\providecommand{\BIBforeignlanguage}[2]{{%
\expandafter\ifx\csname l@#1\endcsname\relax
\typeout{** WARNING: IEEEtran.bst: No hyphenation pattern has been}%
\typeout{** loaded for the language `#1'. Using the pattern for}%
\typeout{** the default language instead.}%
\else
\language=\csname l@#1\endcsname
\fi
#2}}
\providecommand{\BIBdecl}{\relax}
\BIBdecl

\bibitem{sospedra2003use}
M.~Sospedra, C.~Pinilla, and R.~Martin, ``Use of combinatorial peptide libraries for t-cell epitope mapping,'' \emph{Methods}, vol.~29, no.~3, pp. 236--247, 2003.

\bibitem{klinger_multiplex_2015}
\BIBentryALTinterwordspacing
M.~Klinger, F.~Pepin, J.~Wilkins, T.~Asbury, T.~Wittkop, J.~Zheng, M.~Moorhead, and M.~Faham, ``\BIBforeignlanguage{en}{Multiplex {Identification} of {Antigen}-{Specific} {T} {Cell} {Receptors} {Using} a {Combination} of {Immune} {Assays} and {Immune} {Receptor} {Sequencing}},'' \emph{\BIBforeignlanguage{en}{PLOS ONE}}, vol.~10, no.~10, p. e0141561, Oct. 2015, publisher: Public Library of Science. [Online]. Available: \url{https://journals.plos.org/plosone/article?id=10.1371/journal.pone.0141561}
\BIBentrySTDinterwordspacing

\bibitem{nolan_large-scale_2025}
\BIBentryALTinterwordspacing
S.~Nolan, M.~Vignali, M.~Klinger, J.~N. Dines, I.~M. Kaplan, E.~Svejnoha, T.~Craft, K.~Boland, M.~W. Pesesky, R.~M. Gittelman, T.~M. Snyder, C.~J. Gooley, S.~Semprini, C.~Cerchione, F.~Nicolini, M.~Mazza, O.~M. Delmonte, K.~Dobbs, G.~Carreño-Tarragona, S.~Barrio, V.~Sambri, G.~Martinelli, J.~D. Goldman, J.~Heath, L.~D. Notarangelo, J.~Martinez-Lopez, B.~Howie, J.~M. Carlson, and H.~S. Robins, ``\BIBforeignlanguage{English}{A large-scale database of {T}-cell receptor beta ({TCR$\beta$}) sequences and binding associations from natural and synthetic exposure to {SARS}-{CoV}-2},'' \emph{\BIBforeignlanguage{English}{Frontiers in Immunology}}, vol.~16, Jan. 2025, publisher: Frontiers. [Online]. Available: \url{https://www.frontiersin.org/journals/immunology/articles/10.3389/fimmu.2025.1488851/abstract}
\BIBentrySTDinterwordspacing

\bibitem{gao2012cle}
X.~Gao and Y.~Guo, ``Cle peptides in plants: Proteolytic processing, structure-activity relationship, and ligand-receptor interaction f,'' \emph{Journal of integrative plant biology}, vol.~54, no.~10, pp. 738--745, 2012.

\bibitem{gasperini_genome-wide_2019}
\BIBentryALTinterwordspacing
M.~Gasperini, A.~J. Hill, J.~L. McFaline-Figueroa, B.~Martin, S.~Kim, M.~D. Zhang, D.~Jackson, A.~Leith, J.~Schreiber, W.~S. Noble, C.~Trapnell, N.~Ahituv, and J.~Shendure, ``A {Genome}-wide {Framework} for {Mapping} {Gene} {Regulation} via {Cellular} {Genetic} {Screens},'' \emph{Cell}, vol. 176, no.~1, pp. 377--390.e19, Jan. 2019. [Online]. Available: \url{https://www.sciencedirect.com/science/article/pii/S009286741831554X}
\BIBentrySTDinterwordspacing

\bibitem{alda2024mapping}
C.~Alda-Catalinas, X.~Ibarra-Soria, C.~Flouri, J.~E. Gordillo, D.~Cousminer, A.~Hutchinson, B.~Sun, W.~Pembroke, S.~Ullrich, A.~Krejci \emph{et~al.}, ``Mapping the functional impact of non-coding regulatory elements in primary t cells through single-cell crispr screens,'' \emph{Genome Biology}, vol.~25, no.~1, p.~42, 2024.

\bibitem{Du2006PoolingDA}
D.~Z. Du and F.~K. Hwang, ``Pooling designs and nonadaptive group testing: Important tools for dna sequencing,'' in \emph{World Scientific}, 2006.

\bibitem{Du2000CombinatorialGT}
D.~Du and F.~Hwang, \emph{Combinatorial Group Testing and Its Applications}, ser. Applied Mathematics.\hskip 1em plus 0.5em minus 0.4em\relax World Scientific, 2000.

\bibitem{savage_survey_1997}
C.~Savage, ``A survey of combinatorial gray codes,'' pp. 605--629, 1997.

\bibitem{mütze2024combinatorialgraycodesanupdated}
\BIBentryALTinterwordspacing
T.~Mütze, ``Combinatorial gray codes-an updated survey,'' 2024. [Online]. Available: \url{https://arxiv.org/abs/2202.01280}
\BIBentrySTDinterwordspacing

\bibitem{colbourn_group_1999}
C.~J. Colbourn, ``Group testing for consecutive positives,'' \emph{Annals of Combinatorics}, vol.~3, no.~1, pp. 37--41, 1999.

\bibitem{muller_consecutive_2004}
M.~Müller and M.~Jimbo, ``Consecutive positive detectable matrices and group testing for consecutive positives,'' \emph{Discrete Mathematics}, vol. 279, no.~1, pp. 369--381, 2004.

\bibitem{muller_cyclic_2008}
------, ``Cyclic sequences of k-subsets with distinct consecutive unions,'' \emph{Discrete Mathematics}, vol. 308, no.~2, pp. 457--464, 2008.

\bibitem{momihara_constructions_2008}
K.~Momihara and M.~Jimbo, ``Some constructions for block sequences of steiner quadruple systems with error correcting consecutive unions,'' \emph{Journal of Combinatorial Designs}, vol.~16, no.~2, pp. 152--163, 2008.

\bibitem{ge_block_2009}
G.~Ge, Y.~Miao, and X.~Zhang, ``On block sequences of steiner quadruple systems with error correcting consecutive unions,'' \emph{SIAM Journal on Discrete Mathematics}, vol.~23, no.~2, pp. 940--958, 2009.

\bibitem{gad_constant-weight_2011}
E.~Gad, M.~Langberg, M.~Schwartz, and J.~Bruck, ``Constant-weight gray codes for local rank modulation,'' \emph{IEEE Transactions on Information Theory}, vol.~57, no.~11, pp. 7431--7442, 2011.

\bibitem{tang_distance-2_1973}
D.~Tang and C.~Liu, ``Distance-2 cyclic chaining of constant-weight codes,'' \emph{{IEEE} Transactions on Computers}, vol. C-22, no.~2, pp. 176--180, 1973.

\bibitem{mutze_proof_2014}
\BIBentryALTinterwordspacing
T.~Mütze, ``Proof of the middle levels conjecture,'' 2014. [Online]. Available: \url{http://arxiv.org/abs/1404.4442}
\BIBentrySTDinterwordspacing

\bibitem{mutze_book_2024}
------, ``A book proof of the middle levels theorem,'' \emph{Combinatorica}, vol.~44, no.~1, pp. 205--208, 2024.

\end{thebibliography}

\end{document}